# Graphene Spaser Description by Rate Equations

Yu.E. Lozovik[1,2,3,4,a)], I.A. Nechepurenko[3,4], A.V. Dorofeenko[3,4,5,b)], S.P. Merkulova[1]

[1]*Institute for Spectroscopy RAS, 5 Fizicheskaya, Troitsk 142190, Russia*

[2]*Moscow National Research Nuclear University MEPhI, 31 Kashirskoe highway, Moscow 115409, Russia*

[3]*All-Russia Research Institute of Automatics, 22 Suschevskaya, Moscow 127055, Russia*

[4]*Moscow Institute of Physics and Technology, 9 Institutskiy pereulok, Dolgoprudny 141700, Moscow Region, Russia*

[5]*Institute for Theoretical and Applied Electromagnetics RAS, 13 Izhorskaya, Moscow 125412, Russia*

In this paper a surface plasmon polariton laser (spaser), which generates surface plasmons in graphene nanoflake, is considered. The peculiarities of spaser, such as strong material dispersion, require revision of basic laser equations. We provide a full derivation of equations of the spaser dynamics starting from the Maxwell-Bloch equations. Optical Bloch equations and rate equations are obtained and the relation of the equation parameters through the physical ones is given. In the case of graphene realization, the numerical parameter values are estimated.

## I. Introduction

The recent development of plasmonics [1-10] made possible the creation of plasmonic devices analogous to those in classical optics. Theoretical and experimental studies of plasmonic lenses, mirrors, and cavities have been performed [11-16]. Some important benefits of plasmonic devices over optical ones are their subwavelength focusing ability and high field intensity leading to strong field-matter interaction. Surface plasmons are important in the field of surface-enhanced spectroscopy. A high localization of plasmons increases sensitivity of the absorption spectroscopy to molecules located at the surface [17-21]. The latter effect also contributes to surface enhancement of Raman scattering (SERS) [22-25], which has made possible the detection of single molecules [26].

Plasmonics applications are limited by ohmic loss in metal. The use of an active medium has been proposed for loss compensation [27, 28] and amplification [29] of surface plasmon-polaritons (SPPs) propagating along active nanostructures. The amplification can lead to SPP generation [30-34]. Further, it has been understood that a surface plasmon localized at a nanoparticle can also be coherently generated by radiationless excitation [35-39]. The experimental realization of such a system, spaser, was reported by several groups [40-44]. In general, the difference between the SPP generator ("SPP laser") and spaser is vague, so that they are often identified [45]. On the other hand, these two devices can be considered as opposed limiting cases, respectively, of large SPP cavity and small (nanoscopic) system. Beside a rich perspective in applications, plasmonic generators are interesting by themselves as pioneering devices of quantum plasmonics [46, 47].

One of the most promising for plasmonics is graphene.[48-53] Graphene is an extremely thin 2D material [54, 55], which has high carrier mobility [56]. This material supports plasmons, and control of electron Fermi level achievable by doping, for example, by the use of a gate electrode [57], makes it applicable from optical to THz region. The latter frequency region is of high interest as it contains vibrational transitions of molecules. The use of graphene opens up opportunities to create highly sensitive compact THz devices. In near-IR graphene plasmons, the localization factor reaches higher values while losses are lower than in metal plasmons [58]. In spite of the relatively low absorption in graphene, the loss limit of the plasmon mean free path is still a major obstacle for graphene plasmonics applications. As a result, active graphene plasmonics began to develop [59]. In particular, graphene plasmon generators based on the use of a gain medium have been proposed [60-62].

In the present paper, the spaser dynamics equations in the single-mode approximation are derived, and the material dispersion is taken into account. In the case of the graphene nanoflake spaser, the numerical estimation of the equation parameters is given.

**II. Optical Bloch equations for the graphene spaser**

Let us consider a spaser, which consists of a gain medium (quantum dot) and a surface plasmon cavity. This cavity can be realized as a graphene nanoflake. Gain medium has larger coupling to the plasmonic modes than to propagating electromagnetic modes. Therefore, only nonradiative emission of plasmons will be taken into account.

The gain medium is modeled as a two-level system. In this case, the equations for the density matrix elements are transformed to those for the gain medium dipole moment $\mathcal{P}$ and population inversion $\mathcal{N}$. Addition of the Maxwell equation for the electric field $\mathcal{E}$ gives the Maxwell-Bloch equations [63-65]:

$$\nabla \times \nabla \times \mathcal{E} + \frac{1}{c^2}\frac{\partial^2}{\partial t^2}\hat{\varepsilon}(\mathbf{r})\mathcal{E} = -\frac{4\pi}{c^2}C(\mathbf{r})\frac{\partial^2 \langle \mathcal{P} \rangle_\mathbf{r}}{\partial t^2} \quad (1)$$

$$\dot{\mathcal{P}} + \left(\frac{1}{T_2}+i\omega_0\right)\mathcal{P} = -\frac{i}{\hbar}\mathbf{d}_{12}\left(\mathbf{d}_{12}^*\cdot\mathcal{E}\right)\mathcal{N} \quad (2)$$

$$\dot{\mathcal{N}} + \frac{1}{T_1}(\mathcal{N} - \mathcal{N}_0) = -\frac{i}{2\hbar}\left(\mathcal{E}^*\mathcal{P} - \mathcal{E}\mathcal{P}^*\right) \qquad (3)$$

The latter two variables are associated with a single particle of the gain medium, which are distributed with spatial density $C(\mathbf{r})$. Angular brackets with the subscript $\mathbf{r}$, $\langle...\rangle_\mathbf{r}$, mean averaging over gain particles, located in a physically infinitesimal volume at the point $\mathbf{r}$. Asterisk means complex conjugation.

Let us assume a case of single-mode spasing regime. In this case, we will describe the field distribution by that in the lasing mode, $\mathbf{E}(\mathbf{r})$. Furthermore, let us use a slow amplitude approximation using the gain medium transition frequency $\omega_0$ as a reference. With these assumptions, the field is written as

$$\mathcal{E}(\mathbf{r},t) = \mathbf{E}(\mathbf{r})\tilde{e}(t)\exp(-i\omega_0 t) \qquad (4)$$

where $\tilde{e}(t)$ is a slow amplitude. The spatial field profile is determined by the following equation:

$$\nabla \times \nabla \times \mathbf{E} - \frac{\omega_c^2}{c^2}\varepsilon_c(\mathbf{r})\mathbf{E} = 0 \qquad (5)$$

where $\varepsilon_c(\mathbf{r}) = \varepsilon(\omega_c, \mathbf{r})$ is the permittivity distribution (including the graphene conductivity) at the cavity mode eigenfrequency $\omega_c$.

Let us substitute (4) into (1) and multiply scalarly both parts of the obtained equation by $\mathbf{E}^*(\mathbf{r})$. Then, let us find the difference of the latter result and complex conjugate of Eq. (5), which is scalarly multiplied by $\mathbf{E}(\mathbf{r})\tilde{e}(t)\exp(-i\omega_0 t)$. The sequence of the described operations can be sketched as $\mathbf{E}^*(\mathbf{r}) \cdot (A1) - (A5)^* \cdot \mathbf{E}(\mathbf{r})$. The result is

$$\{\mathbf{E}^* \cdot \nabla \times \nabla \times \mathbf{E} - \mathbf{E} \cdot \nabla \times \nabla \times \mathbf{E}^*\}\tilde{e}(t)\exp(-i\omega_0 t) +$$
$$+ (\mathbf{E}\mathbf{E}^*)\left[\frac{1}{c^2}\frac{\partial^2}{\partial t^2}(\hat{\varepsilon}\tilde{e}(t)\exp(-i\omega_0 t)) + \frac{\omega_c^2}{c^2}\varepsilon_c^*\tilde{e}(t)\exp(-i\omega_0 t)\right] = -\frac{4\pi}{c^2}C\mathbf{E}^* \cdot \frac{\partial^2 \langle\mathcal{P}\rangle_\mathbf{r}}{\partial t^2}. \qquad (6)$$

The slow amplitude approximation leads to the following expression for the time derivative in the presence of dispersion (see, Appendix or Refs [66] and [67]):

$$\frac{\partial^2}{\partial t^2}(\hat{\varepsilon}\tilde{e}(t)\exp(-i\omega_0 t)) \approx -\varepsilon_0\omega_0^2\tilde{e}(t)\exp(-i\omega_0 t) - i\frac{\partial(\varepsilon_0\omega_0^2)}{\partial\omega_0}\frac{\partial\tilde{e}}{\partial t}\exp(-i\omega_0 t), \qquad (7)$$

where $\varepsilon_0$ is the permittivity distribution at the transition frequency $\omega_0$. The expression in square brackets in (6) can be expanded as

$$[...] \approx -\varepsilon_0 \frac{\omega_0^2}{c^2} \tilde{e}(t) \exp(-i\omega_0 t) - \frac{i}{c^2} \frac{\partial(\varepsilon_0 \omega_0^2)}{\partial \omega_0} \frac{\partial \tilde{e}}{\partial t} \exp(-i\omega_0 t) + \frac{\omega^2}{c^2} \varepsilon_c^*(z) \tilde{e}(t) \exp(-i\omega_0 t) =$$

$$= -\frac{i}{c^2} \frac{\partial(\varepsilon \omega^2)}{\partial \omega}\bigg|_{\omega_0} \frac{\partial \tilde{e}}{\partial t} \exp(-i\omega_0 t) + \left( \frac{\omega_c^2}{c^2} \varepsilon_c^* - \varepsilon_0 \frac{\omega_0^2}{c^2} \right) \tilde{e}(t) \exp(-i\omega_0 t) =$$

$$= -\frac{i}{c^2} \frac{\partial(\varepsilon \omega^2)}{\partial \omega}\bigg|_{\omega_0} \frac{\partial \tilde{e}}{\partial t} \exp(-i\omega_0 t) + \left( \frac{\omega_c^2}{c^2} (\varepsilon_c' - i\varepsilon_c'') - (\varepsilon_0' + i\varepsilon_0'') \frac{\omega_0^2}{c^2} \right) \tilde{e}(t) \exp(-i\omega_0 t) \approx$$

$$\approx -\frac{i}{c^2} \frac{\partial(\varepsilon \omega^2)}{\partial \omega}\bigg|_{\omega_0} \frac{\partial \tilde{e}}{\partial t} \exp(-i\omega_0 t) + \left( \frac{\delta(\varepsilon' \omega^2)}{c^2} - 2i \frac{\omega_0^2}{c^2} \varepsilon_0'' \right) \tilde{e}(t) \exp(-i\omega_0 t).$$

Here, $\delta(\varepsilon' \omega^2) = \varepsilon_c' \omega_c^2 - \varepsilon_0' \omega_0^2 \approx \frac{\partial(\varepsilon' \omega^2)}{\partial \omega} \delta \omega$ and $\delta \omega = \omega_c - \omega_0$.

Therefore, Eq. (6) transforms to the following form:

$$\{ \mathbf{E}^* \cdot \nabla \times \nabla \times \mathbf{E} - \mathbf{E} \cdot \nabla \times \nabla \times \mathbf{E}^* \} \tilde{e}(t) +$$
$$+ (\mathbf{E}\mathbf{E}^*) \left[ -\frac{i}{c^2} \frac{\partial(\varepsilon \omega^2)}{\partial \omega} \frac{\partial \tilde{e}}{\partial t} + \left( \frac{\partial(\varepsilon' \omega^2)}{\partial \omega} \frac{\delta \omega}{c^2} - 2i \frac{\omega_0^2}{c^2} \varepsilon_0'' \right) \tilde{e}(t) \right] = \qquad (8)$$
$$= -\frac{4\pi}{c^2} C \mathbf{E}^* \cdot \frac{\partial^2 \langle \mathcal{P} \rangle_r}{\partial t^2} \exp(i\omega_0 t).$$

Then, Eq. (8) is integrated over the volume. The expression in curly brackets transforms in the following manner:

$$\int \{ \mathbf{E}^* \cdot \nabla \times \nabla \times \mathbf{E} - \mathbf{E} \cdot \nabla \times \nabla \times \mathbf{E}^* \} dV = \oint \{ \mathbf{E}^* \cdot d\mathbf{S} \times [\nabla \times \mathbf{E}] - \mathbf{E} \cdot d\mathbf{S} \times [\nabla \times \mathbf{E}^*] \} =$$
$$= \oint \{ [\nabla \times \mathbf{E}] \times \mathbf{E}^* - [\nabla \times \mathbf{E}^*] \times \mathbf{E} \} \cdot d\mathbf{S}$$

The latter expression describes the radiation loss, which will be disregarded in comparison with the absorption loss in graphene. Thus, Eq. (8) assumes the form

$$\left[ -iW \frac{\partial \tilde{e}}{\partial t} + \left( W \delta \omega - i \frac{\omega_0}{4\pi} \int \varepsilon_0'' (\mathbf{E}\mathbf{E}^*) dV \right) \tilde{e}(t) \right] = \frac{\omega_0}{2} \int C \left( \mathbf{E}^* \cdot \langle \mathcal{P} \rangle_r \right) \exp(i\omega_0 t) dV. \qquad (9)$$

Here, the notation

$$W = (8\pi\omega_0)^{-1} \int \frac{\partial(\varepsilon'\omega^2)}{\partial\omega}\bigg|_{\omega_0} (\mathbf{EE}^*) dV \qquad (10)$$

is introduced, the imaginary part of the factor $\partial(\varepsilon\omega^2)/\partial\omega$ is neglected, and the time derivative is resolved as $\partial^2 \mathcal{P}/\partial t^2 \approx -\omega_0^2 \mathcal{P}$.

Finally, we introduce the field amplitude $e(t) = \tilde{e}(t)\sqrt{W}$, which does not depend on the normalization of the coordinate function $\mathbf{E}(\mathbf{r})$:

$$\mathcal{E}(\mathbf{r},t) = \mathbf{E}(\mathbf{r}) e(t) \exp(-i\omega_0 t) / \sqrt{W} \qquad (11)$$

Introducing the field relaxation time $T$ as

$$1/T = \frac{\omega_0}{4\pi} \int \varepsilon_0'' (\mathbf{EE}^*) dV / W \qquad (12)$$

and the slow amplitude of the gain medium polarization as

$$p(t) = \int C\left(\mathbf{E}^* \cdot \langle \mathcal{P} \rangle_{\mathbf{r}}\right) \exp(i\omega_0 t) dV / \sqrt{W}, \qquad (13)$$

one writes the field equation:

$$\dot{e}(t) + (i\delta\omega + 1/T) e(t) = i\frac{\omega_0}{2} p(t). \qquad (14)$$

To obtain an equation for $p(t)$, one should average both parts of Eq. (2) over a physically infinitesimal volume ($\langle ... \rangle_{\mathbf{r}}$ operation), then scalarly multiply them by $W^{-1/2} C \mathbf{E}^* \exp(i\omega_0 t)$ and integrate over the volume. This gives

$$\dot{p} + p/T_2 = (i\hbar)^{-1} e(t) W^{-1} \int (\mathbf{d}_{12}\mathbf{E}^*(\mathbf{r}))(\mathbf{d}_{12}^*\mathbf{E}(\mathbf{r})) C(\mathbf{r}) \langle \mathcal{N} \rangle_{\mathbf{r}} dV. \qquad (15)$$

The integral in the latter equation describes the interaction of field with inhomogeneously distributed population inversion $\mathcal{N}(\mathbf{r})$. Let us assume that the effects of gain medium inhomogeneity are not substantial in the system under consideration. Then, one can substitute $\mathcal{N}(\mathbf{r},t)$ by an average value of the population inversion of single particle, which is equal to the ratio of total population inversion of the quantum dots, $D(t)$, to the number of the quantum dots $N$:

$$\langle \mathcal{N}(\mathbf{r},t) \rangle \approx D(t)/N. \qquad (16)$$

Then, the integral in the right-hand side of Eq. (15) transforms to

$$\int (\mathbf{d}_{12}\mathbf{E}^*(\mathbf{r}))(\mathbf{d}_{12}^*\mathbf{E}(\mathbf{r})) C(\mathbf{r}) \mathcal{N}(\mathbf{r}) dV \approx D(t) \int \left| \mathbf{d}_{12}\mathbf{E}^* \right|^2 C(\mathbf{r}) dV / N = D(t) \left\langle \left| \mathbf{d}_{12}\mathbf{E}^* \right|^2 \right\rangle.$$

In this approximation, introducing the interaction parameter $\Xi$ by the relation

$$\hbar \Xi = \left\langle \left| (\mathbf{d}_{12}\mathbf{E}^*) \right|^2 \right\rangle / W, \qquad (17)$$

we obtain the equation for the gain medium polarization:

$$\dot{p} + p/T_2 = -i\Xi D e. \qquad (18)$$

To obtain the equation for the population inversion, let us first average both parts of Eq. (3) over the physically infinitesimal volume (again, $\langle ... \rangle_{\mathbf{r}}$ operation). At the second stage, let us express the field according to Eq. (11) and then multiply both parts of the obtained equation by the concentration $C(\mathbf{r})$ and integrate over the space. Then, taking Eq. (10) into account, we have

$$\int C(\mathbf{r}) \left[ \langle \dot{\mathcal{N}} \rangle + (\langle \mathcal{N} \rangle - \langle \mathcal{N}_0 \rangle)/T_1 \right] dV =$$
$$= -(i/\hbar) \int C(\mathbf{r}) \left[ e^*(t)(\mathbf{E}^* \cdot \langle \mathcal{P} \rangle) \exp(i\omega_0 t) - e(t)(\mathbf{E} \cdot \langle \mathcal{P} \rangle^*) \exp(-i\omega_0 t) \right] dV / \sqrt{W}.$$

According to the above definition, $\int C \langle \mathcal{N} \rangle dV = D(t)$ and, analogously, $\int C \langle \mathcal{N}_0 \rangle dV = D_0$. The latter note, along with the definition of the polarization amplitude (13), brings us to the equation for the population inversion:

$$\dot{D} + (D-D_0)/T_1 = -(i/2\hbar)(e^*p - ep^*). \qquad (19)$$

To summarize, the laser dynamics in the single-mode approximation is described by Eqs. (14), (18), (19):

$$\dot{e}(t) + (i\delta\omega + 1/T)e(t) = i\frac{\omega_0}{2} p(t), \qquad (20)$$

$$\dot{p} + p/T_2 = -i\Xi D(t)e(t), \qquad (21)$$

$$\dot{D} + (D-D_0)/T_1 = -(i/2\hbar)(e^*p - ep^*), \qquad (22)$$

with the parameters given by Eqs. (12), (17).

Let us note that a single particle population inversion cannot exceed a unity in the absolute value. Thus, the total population inversion is $-N \leq D \leq N$.

### III. Rate equations

Despite the plasmonic generator based on noble metal, the one based on graphene can possess so small loss, that the relation between time parameters may be the same as in the case of semiconductor laser: $T_2 \ll T, T_1$. In this case, the gain medium polarization can be substituted by its stationary value

$$p = -i \Xi T_2 D e, \tag{23}$$

which is found from Eq. (21) by setting $\dot{p} = 0$ in the assumption of zero cavity eigenfrequency detuning from the gain medium transition frequency, $\delta\omega = 0$. Then, Eqs. (20) and (22) transform to

$$\dot{e} + e/T = (\omega_0 T_2 / 2) \Xi D e, \tag{24}$$

$$\dot{D} + (D - D_0)/T_1 = -(\Xi T_2 / \hbar) D e e^*. \tag{25}$$

Finally, let us introduce the number of the cavity plasmons as $n(t) = e(t) e^*(t) / (\hbar \omega_0)$. This definition agrees with the relation between the electromagnetic field energy and the number of plasmons: $(8\pi\omega_0)^{-1} \int \left. \dfrac{\partial (\varepsilon' \omega^2)}{\partial \omega} \right|_{\omega_0} (\mathcal{E}\mathcal{E}^*) dV = n\hbar\omega_0$. The equations for $n$ and $D$ are written as

$$\dot{n}(t) + \gamma n(t) = \Omega n(t) D(t), \tag{26}$$

$$\dot{D}(t) + (D(t) - D_0)/T_1 = -\Omega n(t) D(t). \tag{27}$$

where $\Omega = \omega_0 \Xi T_2$ and $\gamma = 2/T$.

### IV. Numerical estimation of the parameters $\Omega$, $\gamma$

According to the above derivation, the parameters in Eqs. (26), (27) are defined as

$$\Omega = \frac{8\pi\omega_0^2 T_2 \int \left|\left(\mathbf{d}_{12}\mathbf{E}^*\right)\right|^2 C(\mathbf{r})dV}{\hbar \int \left.\frac{\partial(\varepsilon'\omega^2)}{\partial\omega}\right|_{\omega_0} (\mathbf{EE}^*)dV}, \qquad (28)$$

$$\gamma = \frac{4\omega_0^2 \int \varepsilon''(\omega_0)(\mathbf{EE}^*)dV}{\int \left.\frac{\partial(\varepsilon'\omega^2)}{\partial\omega}\right|_{\omega_0} (\mathbf{EE}^*)dV}. \qquad (29)$$

In the structure under study, $\varepsilon$ is equal to the permittivity of the substrate at $z<0$, to the surface permittivity of graphene at $z=0$ ($\varepsilon = \delta(z)4\pi i\sigma/\omega$, $\delta(z)$ being the Dirac delta function and $\sigma$ the graphene surface conductivity) and to 1 at $z>0$. Therefore, we have $\varepsilon'' = \delta(z)4\pi\sigma'/\omega$, which reduces the numerator in (29) to $16\pi\omega_0\sigma'|E_t|_g^2 A$, where $A$ is the graphene sensor area, $E_t$ is the tangential component of electric field and the subscript "g" means that the field intensity is calculated at the graphene layer. In the same way, considering the gain layer as being very thin, one can assess the numerator in (28) as $8\pi\omega_0^2 T_2 \left|\left(\mathbf{d}_{12}\mathbf{E}_g^*\right)\right|^2 N$. In the assumption of tangential direction of the gain medium dipole moment, the latter expression transforms to $8\pi\omega_0^2 T_2 |\mathbf{d}_{12}|^2 |E_t|_g^2 N$. The integral in denominators in (28), (29) is calculated in the following manner:

$$\int \left.\frac{\partial(\varepsilon'\omega^2)}{\partial\omega}\right|_{\omega_0} (\mathbf{EE}^*)dV = -4\pi \left.\frac{\partial(\sigma''\omega)}{\partial\omega}\right|_{\omega_0} |E_t|_g^2 A + 2A\omega\int_0^\infty (\mathbf{EE}^*)dz + 2A\omega\varepsilon_D' \int_{-\infty}^0 (\mathbf{EE}^*)dz,$$

where $\varepsilon_D$ is permittivity of the dielectric substrate. Let $\kappa$ and $\kappa_D$ be the imaginary parts of the normal components of surface plasmon wave numbers in vacuum and dielectric, so that $\kappa^2 = k^2 - k_0^2$ and $\kappa_D^2 = k^2 - k_0^2\varepsilon_D$, where $k$ is the surface plasmon wave number and $k_0 = \omega/c$. The z dependence of electric field is $(\mathbf{EE}^*) = |\mathbf{E}|_g^2 \exp(2\kappa_D z)$ in dielectric ($z<0$) and $(\mathbf{EE}^*) = |\mathbf{E}|_g^2 \exp(-2\kappa z)$ ($z>0$). This leads to the following expression for the integral in denominators in (28), (29):

$$\int \left.\frac{\partial(\varepsilon'\omega^2)}{\partial\omega}\right|_{\omega_0} (\mathbf{EE}^*)dV = -4\pi \left.\frac{\partial(\sigma''\omega)}{\partial\omega}\right|_{\omega_0} |E_t|_g^2 A + 2A\omega|\mathbf{E}|_g^2/(2\kappa) + 2A\omega\varepsilon_D'|\mathbf{E}|_g^2/(2\kappa_D).$$

Finally, let us note the relation between the tangential and normal electric field components:

$kE_t + i\kappa E_n = 0$ in vacuum and $kE_t + i\kappa_D E_n = 0$ in dielectric. Along with the relation $|\mathbf{E}|^2 = |E_t|^2 + |E_n|^2$, we find $|\mathbf{E}|^2_g = |E_t|^2_g \frac{2k^2 - k_0^2}{k^2 - k_0^2}$ in vacuum and $|\mathbf{E}|^2_g = |E_t|^2_g \frac{2k^2 - k_0^2 \varepsilon_D}{k^2 - k_0^2 \varepsilon_D}$ in dielectric. For the graphene plasmon, $k^2 \gg k_0^2$, which means $|\mathbf{E}|^2_g \approx 2|E_t|^2_g$. In the same approximation, $\kappa^2 \approx \kappa_D^2 \approx k^2$, which leads to

$$\int \left.\frac{\partial(\varepsilon'\omega^2)}{\partial\omega}\right|_{\omega_0} (\mathbf{E}\mathbf{E}^*)dV = \left[-4\pi \left.\frac{\partial(\sigma''\omega)}{\partial\omega}\right|_{\omega_0} + 4\omega/k\right]|E_t|^2_g A.$$

The expressions for the constants transform to

$$\Omega = \frac{2\pi\omega_0^2 T_2 |\mathbf{d}_{12}|^2 C_S}{\hbar \left[-\pi \partial(\sigma''\omega)/\partial\omega\big|_{\omega_0} + \omega_0/k\right]}, \quad (30)$$

$$\gamma = \frac{4\pi\omega_0 \sigma'}{\left[-\pi \partial(\sigma''\omega)/\partial\omega\big|_{\omega_0} + \omega_0/k\right]}, \quad (31)$$

where $C_S = N/A$ means the surface density of the quantum dots.

For the frequency $\omega_0$ corresponding to the wavelength of 6 mkm the graphene conductivity is $\sigma' = 1.68 \cdot 10^6$ cm/s and $\partial(\sigma''\omega)/\partial\omega\big|_{\omega_0} = 16759$ cm/s and the graphene plasmon wavenumber is $k = 239827 + 1192i$ cm$^{-1}$. We consider the quantum dot parameters $T_2 = 30$ fs, $d_{12} = 20$ D (1 D = $10^{-18}$ CGS unites) and quantum dot concentration corresponding to 20 nm x 20 nm area for each quantum dot. A spacer, which decreases the SPP local field intensity by $10^{-2}$ between the graphene sheet and the gain medium is supposed. This gives $\gamma = 1.56 \times 10^{12} s^{-1}$ and $\Omega = 5.3 \times 10^9 s^{-1}$.

**Conclusions**

In conclusion, we have obtained the expressions for the parameters, which are needed to describe the spaser dynamics. The numerical values of these parameters in the case of graphene spaser are estimated.

**Appendix**

In this Appendix we provide a derivation of the relation

$$\frac{\partial^2}{\partial t^2}\hat{\varepsilon}\left[e(t)\exp(-i\omega_0 t)\right] = \left[-\omega_0^2 \varepsilon(\omega_0) e(t) - i\frac{\partial(\varepsilon\omega^2)}{\partial\omega}\dot{e}(t)\right]\exp(-i\omega_0 t),$$

with $e(t)$ being a slow amplitude.

A causal time operator $\hat{\varepsilon}$ with the kernel $G(\tau)$ is defined as

$$\hat{\varepsilon}E(t) = \int_0^\infty d\tau\, G(\tau) E(t-\tau). \tag{A1}$$

Let us represent the field through the slow amplitude, $E(t) = e(t)\exp(-i\omega_0 t)$, and calculate the time amplitude. The second derivative of the slow amplitude is disregarded:

$$\frac{\partial^2}{\partial t^2}\hat{\varepsilon}\left[e(t)\exp(-i\omega_0 t)\right] = \frac{\partial^2}{\partial t^2}\left[\exp(-i\omega_0 t)\int_0^\infty d\tau\, G(\tau) e(t-\tau)\exp(i\omega_0\tau)\right] \approx$$

$$\approx \left[-\omega_0^2 \int_0^\infty d\tau\, G(\tau) e(t-\tau)\exp(i\omega_0\tau) - 2i\omega_0 \int_0^\infty d\tau\, G(\tau) \dot{e}(t-\tau)\exp(i\omega_0\tau)\right]\exp(-i\omega_0 t).$$

Here, dot means the time derivative. The first term will be approximated as $e(t-\tau) \approx e(t) - \tau\dot{e}(t)$, which is justified by a rapid decay of the kernel $G(\tau)$. In the second term, the zero-order approximation is enough: $\dot{e}(t-\tau) \approx \dot{e}(t)$. The integrals, which appear after the described transformations, are easily interpreted. First, after the Fourier transform of both parts of Eq. (A1), one obtains the frequency representation of the permittivity: $\varepsilon(\omega) = \int_0^\infty d\tau\, G(\tau)\exp(i\omega\tau)$. Second, it follows from the latter relation that

$$\partial\varepsilon(\omega)/\partial\omega = i\int_0^\infty \tau d\tau\, G(\tau)\exp(i\omega\tau).$$

After the transformations discussed above, one obtains the following:

$$\frac{\partial^2}{\partial t^2}\hat{\varepsilon}\left[e(t)\exp(-i\omega_0 t)\right] = \left[-\omega_0^2 \varepsilon(\omega_0)e(t) - i\omega_0^2 \left.\frac{\partial \varepsilon(\omega)}{\partial \omega}\right|_{\omega=\omega_0} \dot{e}(t) - 2i\omega_0 \varepsilon(\omega_0)\dot{e}(t)\right]\exp(-i\omega_0 t),$$

which gives the required relation.